\documentclass[prl,preprint,superscriptaddress]{revtex4-1}

\usepackage{geometry}                
\usepackage{graphicx}
\usepackage{amsmath}
\usepackage{amssymb}
\usepackage[outdir=./]{epstopdf}
\usepackage{hyperref}

\hypersetup{
    colorlinks,
    citecolor=black,
    filecolor=black,
    linkcolor=red,
    urlcolor=black
}

\begin{document}

\title{Interacting atomic interferometry for rotation sensing approaching the Heisenberg Limit}

\date{\today}
\author{Stephen Ragole}
\email{ragole@umd.edu}
\affiliation{Joint Quantum Institute, University of Maryland, \\ College Park, MD 20742, USA}
\affiliation{Joint Center for Quantum Information and Computer Science, University of Maryland, \\ College Park, MD 20742, USA}
\author{Jacob M. Taylor}
\affiliation{Joint Quantum Institute, University of Maryland, \\ College Park, MD 20742, USA}
\affiliation{Joint Center for Quantum Information and Computer Science, University of Maryland, \\ College Park, MD 20742, USA}
\affiliation{National Institute of Standards and Technology, \\ Gaithersburg, MD 20899, USA}

\begin{abstract}
Atom interferometers provide exquisite measurements of the properties of non-inertial frames. While atomic interactions are typically detrimental to good sensing, efforts to harness entanglement to improve sensitivity remain tantalizing. Here we explore the role of interactions in an analogy between atomic gyroscopes and SQUIDs, motivated by recent experiments realizing ring shaped traps for ultracold atoms. We explore the one-dimensional limit of these ring systems with a moving weak barrier, such as that provided by a blue-detuned laser beam. In this limit, we employ Luttinger liquid theory and find an analogy with the superconducting phase-slip qubit, in which the topological charge associated with persistent currents can be put into superposition. In particular, we find that strongly-interacting atoms in such a system could be used for precision rotation sensing. We compare the performance of this new sensor to an equivalent non-interacting atom interferometer, and find improvements in sensitivity and bandwidth beyond the atomic shot-noise limit.
\end{abstract}

\maketitle

Cold atomic systems have provided an exciting arena for studying aspects of quantum mechanics. The ability to coherently manipulate atoms has been used to great effect in measuring the properties of non-inertial frames, e.g. \cite{Gustavson1997a,Snadden1998}. In contrast to the usual atomic systems, the strong interactions in superconducting quantum interference devices (SQUIDs), in addition to their coherence and non-linear electrical response, have made them a tool for both making qubits and performing incredibly high precision sensing, e.g. \cite{Orlando2002,Friedman2000a,Wellstood1984,Yurke1988,Siddiqi2004,clarke2006squid}. These applications have also been explored in superfluid helium, where a constriction in superfluid flow acts like a Josephson junction, enabling sensing of rotation, the analogue to magnetic field for neutral particles \cite{Packard1992a}. Employing the superfluid flow of interacting cold atoms, recent experiments creating ring shaped traps have made a connection to SQUIDs directly accessible \cite{Gupta2005,Ryu2007,Ramanathan2011a,Wright2013,Eckel2014c,Moulder2012a,Ryu2013a}. These ring traps allow experimentalists to manipulate atoms in the ring with potential barriers of variable strength and location. We will consider a system where a weak barrier can combine with strong atomic interactions to make an interaction-enhanced gyroscope that employs correlations of many-body excitations, i.e. persistent currents.

Previous approaches to atom interferometric sensing use the ability to transform phase evolution along different paths into population differences, but treat atomic interactions as deleterious to sensitivity \cite{Dimopoulos2008,Jannin2015}.  In these approaches, e.g., Ramsey interferometry, single atoms are put into superposition and the relative phase gained over some time contains information about the quantity to be measured.  The far end of the interferometer converts these phases into measurable population differences. However, atoms can interact during this process, altering the phase and leading to a loss of single atom coherence, which decreases the final sensitivity of the measurement \cite{Dimopoulos2008,Jannin2015}. Usually, these experiments can be engineered to minimize the possibility of interaction and they have produced remarkably precise measurements of gravitation and rotation \cite{Gustavson1997a,Snadden1998}. This precision comes in part from being able to conduct a large number of independent single atom measurements simultaneously.  These ensemble measurements have a noise/signal ratio limited by the shot noise of N independent two-level systems. This noise/signal ratio goes as $\frac{1}{\sqrt{N}}$, commonly known as the shot noise limit \cite{Caves1980}.  Sensitivities may be improved even to the limit from Heisenberg uncertainty, but only through atomic entanglement, such as squeezing \cite{Giovannetti2004}.

This letter describes a system designed to explore the effect of atomic interactions on the sensitivity of an atomic interferometer to rotational flux. In particular, we investigate whether there are situations in which the atomic interactions can lead to the correlations necessary to beat the shot noise limit while not being too strongly dephased to prevent sensitivity improvements. We find that a strongly (but not too strongly) repulsive gas of atoms with a weak barrier can be manipulated to create persistent current-state superpositions, which can be used to sense rotation with sensitivity that scales as $N^{-3/4}$, below the shot noise limit, but not approaching the Heisenberg limit.

Strongly interacting systems are famously challenging.  To reduce the difficulty of this problem, we study only the long wavelength behavior of a gas of atoms trapped in a ring geometry in the 1D limit. While current experiments are, at best, quasi-2D \cite{Ramanathan2011a,Wright2013,Eckel2014c,Moulder2012a,Ryu2013a}, there are no fundamental obstructions to the creation of 1D systems as described above and by others \cite{Citro2009a,Didier2009a, Hallwood2010b,Solenov2010c,Solenov2010b}, with efforts in progress \cite{Cai2015}. This dimensional reduction comes with a great simplification of the physics involved and allows us to consider a variety of interactions and even statistics, though we focus on the bosonic case here.

Since we want to consider a wide range of atomic interactions, perturbative methods may not be suitable. Mean-field approximations, such as those underlying the Gross-Pitaevskii equation, miss a crucial quantum effect: the ability to create superpositions of many-body excitations, which we find to be necessary for interaction-assisted metrological benefit. Instead, we will employ Luttinger liquid theory, an effective field theory which universally describes quantum systems in 1D with short-range interactions \cite{Haldane1981c,Giamarchi2003,Cazalilla2004}.

Specifically, we require temperatures and time variations that are slow compared to the Luttinger energy scale, $E_{LL} \sim \frac{\hbar^2 \pi^2 \rho_0^2}{K}$, where $\rho_0=\langle \rho \rangle$ is the average number of atoms per length. $K$ is the Luttinger parameter, which encodes the combined effects of statistics and interactions. For example, $K=1$ corresponds to the Tonks-Girardeau gas (or free fermions) and $K\rightarrow \infty$ is the superfluid limit.  Notably, there exists a mapping from the interaction strength of delta function-interacting bosons to the Luttinger parameter which will allow us to consider the range of interactions for a repulsive Bose gas \cite{Cazalilla2004}. We will show that having $K\sim 1$ is ideal for the gyroscope. In this limit, we can express the Luttinger parameter in terms of the 3D scattering length, $a_s$, the transverse confinement, $l_{\perp}$, and $\rho_0$. Explicitly

 \begin{align}
 K= 1 + \frac{2 \rho_0 l_{\perp}^2\left(1-\frac{Ca_s}{l_{\perp}}\right)}{a_s}
 \end{align}

 where $C\approx1.0325...$ is a constant \cite{Cazalilla2004}. This theory has the following free Hamiltonian (following conventions from \cite{Cazalilla2004}).
\begin{align}
H_0 &= \frac{\hbar v_s}{2 \pi} \int_0^L \mathrm{d}x \left[ K \left(\partial_x \phi(x)\right)^2 + \frac{1}{K} \left(\partial_x \Theta(x)-\pi \rho_0 \right)^2 \right]
\end{align}
Here, $v_s$ is the speed of sound, $L$ is the circumference of the ring, $\phi(x)$ is a local phase of the underlying field which we are abstracting away. $\partial_x \Theta(x)$ is related to the number density, $\rho$, by $\rho(x)=\frac{\partial_x \Theta(x)}{\pi}\sum_{l=-\infty}^{+\infty}e^{2il\Theta(x)}$.  The Luttinger fields $\phi(x)$ and $\Theta(x)$ have the following commutation relation \begin{align}
[\phi(x),\frac{\partial_{x'} \Theta(x')}{\pi} ] = i \delta (x-x') \end{align}

\begin{figure}
	\includegraphics[width=0.5\textwidth]{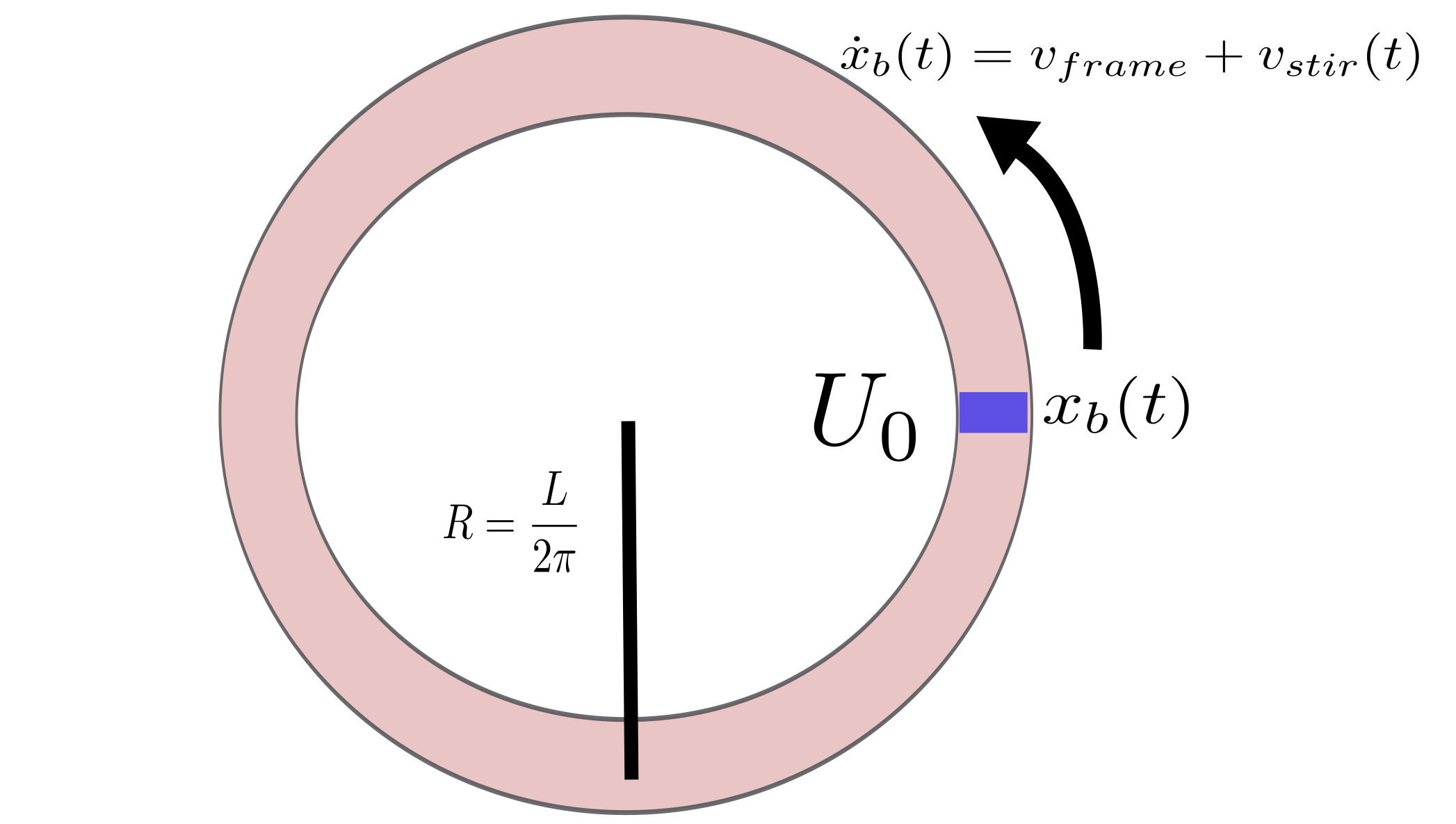}
	\caption{A cartoon of the system. Atoms are trapped in a 1D ring of length $L$ with a blue-detuned laser crossing at a single point, $x_b(t)$. In the pictured atom frame, the barrier rotates through the ring at a rate which is the sum of the laboratory frame rotation ($\omega_{frame} = \frac{2 \pi v_{frame}}{L})$ and the externally controlled stirring rate ($\omega_{stir} = \frac{2 \pi v_{stir}}{L})$.}
	\label{fig:cartoon}
\end{figure}

To make the system sensitive to rotation, we break rotational symmetry by adding a blue-detuned laser beam as a localized potential barrier shown in Fig. \ref{fig:cartoon}.  We approximate the laser in the long wavelength theory as a (moving) barrier at a single point ($x=x_b(t)$) on the ring. When the barrier is smaller than $E_{LL}$, i.e., a weak barrier, it induces a new term in the Hamiltonian \cite{Cazalilla2004,Giamarchi2003, Kane1992c,Kane1992b}.
\begin{align}
V&=\nonumber \int_0^L \mathrm{d}x U_0 \delta(x-x_b(t)) \rho(x) \\
&\approx 2 N U_0 \cos(2\Theta(x_b(t)))
\end{align}
$N$ is the particle number, and $U_0$ is the dipole potential from the laser. In this expansion, we have only kept the lowest harmonics of the density (consistent with \cite{Giamarchi2003, Kane1992c,Kane1992b}). Though strong laser barriers have been used to create traps with `weak-links,' the weak barrier considered here preserves the character of persistent current states and allows us to couple them perturbatively. As topological excitations, these persistent currents will be visible in time-of-flight images as vortices.

We perform a standard field expansion for periodic boundary conditions
\cite{Cazalilla2004, Cominotti2014d}. These boundary conditions assume Galilean invariance which insures that phonons, $b_q$, and topological excitations will be uncoupled in the absence of a barrier. In this expansion,

\begin{align}
H_0 &= \sum_{q\ne 0} \hbar \omega(q) b^{\dagger}_q b_q + \frac{\rho_0 L\hbar \omega_0}{8 K^2}(N-\rho_0 L)^2 + \frac{\rho_0 L\hbar \omega_0}{2} J^2
\end{align}
$\omega_0 =\frac{4 \pi^2 \hbar}{ML^2}$ is the rotation quantum for particles of mass $M$ in a ring of circumference $L$  and we use the fact that Galilean invariance gives us $v_s K =\frac{\hbar \pi N}{ML}$ to achieve this form. This substitution allows us to rewrite the Hamiltonian entirely in terms of the phonon energies and the fundamental energy scale of the ring, $\hbar \omega_0$. We restrict our consideration to a fixed atom number ($N=\rho_0 L$). The current operator, $J$, denotes phase winding, i.e. $\phi(x) \propto \frac{2 \pi x}{L} J$. As a topological quantity, $J$ has integer eigenvalues and represents the topological charge associated with persistent current in the ring. The phonon modes, $b_q$, are bosons with quasimomentum $q_n = \frac{2 \pi  n}{L}$ for  $n\in \mathbb{Z}$ and $\omega(q) = \hbar v_s |q|$ for $q \ll \rho_0.$  

Now, we transform to a frame which is co-rotating with the barrier. The barrier both rotates along with the lab frame and can be actively controlled relative to the lab frame to `stir' the gas. Specifically, control of the barrier rotation rate can be used to engineer topological charge superpositions which will be essential for interaction-assisted sensing. We now proceed with the transformation. In the field expansion used above, $\Theta(x) = \theta_0 + \frac{\pi x}{L}N +\sum_{q\ne0} \left|\frac{2 \pi K}{qL}\right| (e^{iqx}b_q +e^{-iqx}b_q^{\dagger})$ \cite{Cazalilla2004}. $\theta_0$ is the zero-mode of the field.  Noting $[J,2\theta_0] = i$ and $[b_q,b_{q'}^{\dagger}] = \delta_{qq'}$, we transform the Hamiltonian with $U_{\textrm{RF}} = \mathrm{exp}\left(-i\left[\frac{2\pi x_{b}(t)N}{L}J+\sum_{q\ne0}  q x_{b}(t) b_q^{\dagger}b_q\right]\right)$.

The free Hamiltonian is invariant under the transformation, while $V \rightarrow 2 N U_0 \cos(2\Theta(0))$.  We also gain the terms

\begin{align}
\delta H = -i\hbar U^\dagger_{\textrm{rf}} \dot{U}_{\textrm{rf}} = -\hbar \omega_b(t) NJ - \hbar \sum_{q\ne0} q \dot{x}_b(t)b^{\dagger}_q b_q
\end{align}
where $\omega_b(t)$ is the angular frequency of the barrier rotation relative to the atoms.

\begin{figure}
	\includegraphics[width=0.5\textwidth]{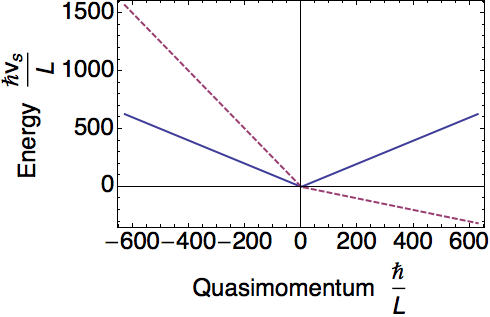}
	\caption{The energy of long wavelength phonons in a non-rotating system (blue, solid) and phonons in a system rotating supersonically with $\dot{x}_b =1.5 v_s$ (red, dashed). In the rotating system, the forward traveling phonons have negative energy, signaling instability.}
	\label{fig:Phonon}
\end{figure}

We complete the square for the linear $J$ term and ignore the constant term produced under our fixed atom number assumption.  Thus, the transformation leads to a shift in the persistent current operator $J^2 \rightarrow  \left(J- \frac{\omega_b(t)}{\omega_0}\right)^2$ where $\omega_b = \frac{2\pi \dot{x}_b(t)}{L}$.  The phonon term is easily absorbed by defining a new phonon dispersion relation, $\tilde{\omega}(q) = v_s |q| - \dot{x}_b(t) q$. This shifted frequency confirms the intuition that if the stirring speed, $\dot{x}_b$, grows larger than the sound velocity, $v_s$, our theory will become unstable, pictured in Fig. \ref{fig:Phonon}.

Adding in a barrier breaks the Galilean invariance, and can couple phonons and topological excitations to themselves and each other, potentially decohering topological charge superpositions. We can expand the barrier term using the field expansion above
\begin{align}
V &= 2 N U_0 \cos(2\Theta(0)) \\ \nonumber
 &= N U_0\bigg(e^{2i(\theta_0 + \delta\theta(0))} + e^{-2i(\theta_0 + \delta\theta(0))}\bigg)
\end{align}
where $\delta\theta(0)= \sum_{q\ne0} |\frac{2 \pi K}{qL}| (b_q +b^{\dagger}_q)$ is the phonon contribution to the field.

We focus on the coupling of topological charge states, which are most suited for sensing applications. Thus, we integrate over the phonon modes to determine the effective interaction between the persistent current states. Following \cite{Kane1992c,Kane1992b,Cominotti2014d}, we arrive at the following expression for the potential barrier.
\begin{align}
\label{potential}
V&= N U_0 e^{2i\theta_0} \langle e^{2i\delta\theta(0)}\rangle_{\delta\theta} +h.c. \\ \nonumber
&= 2 N U_{\textrm{eff}} \cos(2\theta_0)
\end{align}
where the brackets denote functional integration over the phonon modes and $U_{\textrm{eff}} = U_0(\frac{d}{L})^K$ is the renormalized barrier strength, and $d$ is a short distance cutoff. While Luttinger liquid theory has a cutoff above which it loses validity ($E_{LL} \approx \frac{N^2 \hbar \omega_0 }{4K}$), this renormalization step will give a lower cutoff, $E_{ph} =\frac{N\hbar \omega_0 }{4K} \approx \frac{E_{LL}}{N}$.  This new cutoff generates a timescale below which the renormalized theory is not valid, which will be important to consider when manipulating the system. Simply put, working below the lowest phonon mode frequency prevents decoherence but dramatically reduces the effective barrier height and also lowers the ``max velocity" for diabatic processes. 

Integrating out the phonons renormalizes the barrier in a manner that depends on both the microscopic details and the Luttinger parameter, $K$. Here we see the first non-trivial indication of the interactions: in the superfluid limit ($K\rightarrow \infty$), a barrier will be weakened significantly by the phononic modes. However, in the strongly repulsive ($K\rightarrow 1$) regime, the barrier will remain finite, allowing mixing between current states. In the strongly repulsive regime, the relevant cut-off is $d\approx\frac{K L}{N}$, so $U_{\textrm{eff}} = U_0(\frac{K}{N})^K$ \cite{Cominotti2014d}.

This Hamiltonian is similar to the quantum phase slip junction, e.g. \cite{Mooij2006}, and is the dual of the standard superconducting charge qubit Hamiltonian \cite{Shnirman1997}:
\begin{align}
H_{JJ} = E_c (n-n_g)^2 -E_J\cos(\delta)
\end{align}
where $n$ is the number of Cooper pairs on the island, $n_g$ is set by the gate voltage and $\delta$ is the phase difference across the junction and $[\delta, n] = i$. Under the transformation $n \rightarrow J$ and $\delta \rightarrow -2\theta_0$, the atoms in the ring form a charge-qubit-like system with $E_C = \frac{N\hbar \omega_0}{2} $ and $E_J = 2 N^{1-K} U_0 K^K$ ($E_C \gg E_J$, since the barrier is perturbative).

\begin{figure}
	\includegraphics[width=1.0\textwidth]{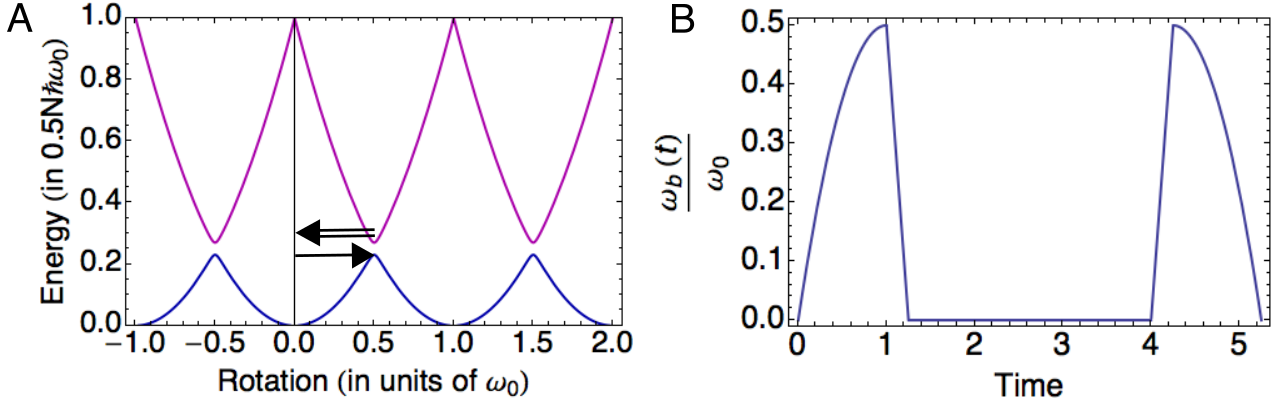}
	\caption{A. The energy spectrum for the perturbed current states, blue (magenta) represents the ground (excited) state of the system. The weak barrier creates avoided crossings at rotation equal to odd half-integer rotation quanta. The arrows represent the proposed `$\frac{\pi}{2}$ pulse:' the system is adiabatically driven to the avoided crossing (single arrow) and diabatically returned $\omega=0$ (double arrow). B. A cartoon of the proposed Ramsey sequence. The sequence consists two $\frac{\pi}{2}$ pulses with an observation time $\tau_{\textrm{obs}}$ in between.}
	\label{fig:energies}
\end{figure}

Since the barrier couples the current state $|J\rangle$ to states $|J'\rangle=|J\pm1\rangle$, superpositions can be formed by precisely controlling the rotation rate of the stirring beam. Consider the case of preparing the atoms without any rotation, $|\Psi\rangle = |0\rangle$.  Here, only the states $|\pm1\rangle$ will be coupled by the barrier and only mix very weakly into the ground state at $\omega=0$.  We can implement a `$\frac{\pi}{2}$-pulse' in two steps as illustrated in Fig. \ref{fig:energies}a. First, we adiabatically increase rotation to $\omega=\frac{\omega_0}{2}$, where the instantaneous ground state is $\frac{1}{\sqrt{2}} (|0\rangle -|1\rangle)$.  Then, the rotation rate is diabatically ramped back to $\omega=0$ and barrier turned off.  This process will be completed in a time $\tau_{\pi/2} = \tau_{\textrm{adiabatic}} + \tau_{\textrm{diabatic}}$. These times can be determined from a Landau-Zener analysis of the effective two-level system. Note that each of these times must be longer than the timescale of the renormalized theory, $\tau_{ph} = \frac{4K}{\omega_0 N}$ to prevent phonon-based dephasing.  The adiabatic and diabatic timescales will be set by the effective barrier strength, $\tau_{\pi/2} \propto \frac{\hbar}{NU_{\textrm{eff}}} \gg \tau_{ph}$ which will limit the strength the barrier can take.

Having established the ``charge" qubit-like behavior and appropriate sequences for preparing topological charge superpositions, we now propose a Ramsey interferometry scheme for rotation sensing, using the persistent current states as the basis. As described above, we will create a superposition of current states, $\frac{1}{\sqrt{2}} (|0\rangle -|1\rangle)$, and turn the barrier off. Then, we will expose this superposition to a small rotation rate ($\omega \ll \omega_0$) for a time $\tau_{\textrm{obs}}$ without the barrier.  Over this time, the superposition will evolve into the state $\frac{1}{\sqrt{2}} (|0\rangle -e^{i\phi}|1\rangle)$ where $$\phi = N \tau_{\textrm{obs}} \Big(\frac{\omega_0}{2}-\omega\Big)$$  This phase can be converted into a population difference by turning the barrier back on and performing another `$\frac{\pi}{2}$ pulse.' A cartoon of the process is pictured in Fig. \ref{fig:energies}b. The final state then can be projectively measured by the vortex detection techniques, e.g., using time-of-flight \cite{Wright2013}. Since the phase scales with the number of atoms while the vortex shot noise is constant, the nominal sensitivity to rotation has Heisenberg-like scaling in the absence of noise.

Here we pause to note the crucial role of interactions in creating these states. In the non-interacting ($K\rightarrow \infty$) limit, the gap between current states closes much more rapidly as a function of system size, making the adiabatic process needed for superposition unfeasible. Therefore, the proposed gyroscope is most viable in the strongly interacting limit which maintains the gap needed to couple persistent current states. While our analysis has only considered a perfectly clean system, it is likely that there will be some disorder present in the trap. Given the strength of fluctuations in 1D, this disorder leads to localization for $K<1.5$ \cite{Giamarchi2003}.  Therefore, the optimal $K$ is just above this localization limit.

To compare with other gyroscopes, we must consider realistic sources of noise that could affect the sensitivity. In particular, we will consider noise in the atom number, which will vary from shot-to-shot.  This variation will give different output phases with a constant rotation signal, so it must be understood to quantify how sensitive the sensor can be. Other systematic noise issues, such as laser power and trap configuration fluctuations, will be problematic but can be surmounted with sufficient detuning and laser power.

To compute the effect of the shot-to-shot variations in atom number, we assume that number fluctuations are Poissonian, $\sigma_N = \sqrt{N}$.  For large $N$, we can approximate the Poisson distribution as a Gaussian, centered at $N$ with $\sigma_N = \sqrt{N}$. In this approximation, we can consider each individual run of the experiment as having some fixed signal and a random additional noise.  For convenience, we define $F =\frac{\omega_0}{2}-\omega$.  The phase of a given experiment is:
\begin{align}
\nonumber \phi_i &= \phi_0 + \phi_{\delta_i} \\
&= N F\tau+ \delta N_i F \tau
\end{align}
Now, we consider an average of many measurements over the noise.
\begin{align}
\nonumber \langle e^{i \phi} \rangle &= \langle e^{i (NF\tau+ \delta N_i F \tau)} \rangle \\
\nonumber &= e^{iN F\tau} \langle e^{i\delta N_i F \tau} \rangle \\
& = e^{i N F\tau} e^{- F^2 \sigma_N^2\tau^2}
\end{align}
where, again, $\sigma_N$ is the standard deviation of the atom number.

While in the absence of noise, longer evolution times would result in higher sensitivity, the low frequency noise decreases contrast as $e^{-F^2 \sigma_N^2 \tau^2}$ as $\tau$ increases.  With noise added in, we can calculate the sensitivity of our experiment.

\begin{align}
S &= \bigg|\partial_\omega \frac{\textrm{Signal}}{\textrm{Noise}} \bigg|^{-1}\bigg|_{\omega =0} \sqrt{\tau_{\textrm{obs}}} \\ \nonumber
 &= \bigg|  \frac{N \omega_0 \tau_{\textrm{obs}}\sin(N F_0 \tau_{\textrm{obs}}+\phi_0)e^{-F_0^2 \sigma_N^2\tau_{\textrm{obs}}^2}}{2F_0}\bigg|^{-1} \sqrt{\tau_{\textrm{obs}}} \\ \nonumber
& = \frac{2 F_0e^{F_0^2 \sigma_N^2\tau_{\textrm{obs}}^2}}{N \omega_0 \sqrt{\tau_{\textrm{obs}}}} \\ \nonumber
& = \frac{e^{\frac{\omega_0^2}{4} \sigma_N^2\tau_{\textrm{obs}}^2}}{N \sqrt{\tau_{\textrm{obs}}}}
\end{align}
where in the last line, we have used $F_0 = F(\omega=0) = \frac{\omega_0}{2}$.

From this expression, we can optimize $\tau_{\textrm{obs}}$ and determine the maximum sensitivity for the device. Using the optimum observation time, $\tau^*_{\textrm{obs}} = \frac{1}{\omega_0 \sigma_N}$, we calculate the optimal sensitivity.

\begin{align}
S_{\textrm{max}}& = \frac{e^{1/4} \sqrt{\omega_0 \sigma_N} }{N}
\end{align}

Here we see that under the assumption that $\sigma_N= \sqrt{N}$, the nominal Heisenberg-like scaling for $N$ fixed is converted into $N^{-3/4}$ scaling. However, this is an improvement over the shot-noise limit and could be further enhanced if the time-of-flight images from vortex detection are calibrated to give an estimate of atom number. With an estimate of atom number, the rotation can be more precisely estimated by fitting the slope of pairs of vortex number (0 or 1) and estimated atom number. For small signals, these data will form a line and the noise propagation is straightforward.

Using the experimental temperature of $100$ nK and ring radius $R=19.2 \mu$m \cite{Ramanathan2011a,Wright2013,Eckel2014c}, we assume a transverse confinement of $l_{\perp} \approx 200 $nm, which gives $a_s \approx 3600 a_0$, where $a_0$ is the Bohr radius, to set $K\approx 1.6$ for $N=10^5$. Estimating $\sigma_N = \frac{\sqrt{N}}{10}$, we find that a sensor with $N=10^5$ atoms would have $\tau^*_{\textrm{obs}} \approx 4$ ms, a sensitivity of $~2 \times10^{-4} \frac{\textrm{rad}}{s\sqrt{Hz}}$ and a bandwidth $ \geq 200$ Hz. Since the entanglement allows relatively rapid phase accumulation, the sensor has a higher bandwidth than single-atom based sensors. For single-shot readout, such short free evolution times make non-interacting atom interferometry challenging. To reasonably compare, we instead consider a sensitivity per root bandwidth.

We can plot the numerical results for optimum sensitivity as a function of atom number, $N$, and compare with the noiseless limit and an atom interferometer as described in \cite{Gustavson1997a}, each evaluated for a fixed time $\tau_{\textrm{comp}}=\frac{2\pi}{\omega_0}=0.838$ s. This time is set by travel time for atoms moving at the persistent current velocity to circumnavigate the ring. This time is much longer than optimal observation for the Luttinger system, $\tau_{\textrm{comp}} \approx 6\times \mathrm{max}(\tau_{\textrm{obs}}^*)$. In the atom interferometer, the atoms will gain a Sagnac phase of $\phi = \frac{2M}{\hbar} \omega A$ where $A = \frac{L^2}{4\pi^2}$ is the area enclosed by the atoms.  This phase can be conveniently rewritten in terms of $\tau_{\textrm{comp}}$, $\phi = \frac{2\pi M R^2}{\hbar} \omega = \omega \tau_{\textrm{comp}}$. The sensitivity for the comparison single atom system will be

\begin{align}
S_{SA} &= \frac{1}{|\partial_{\omega} (\frac{1}{2} \cos(\omega \tau_{\textrm{comp}}))|} \sqrt{\frac{\tau_{\textrm{comp}}}{N}} \\ \nonumber
S_{SA_{max}} &= \frac{2}{\sqrt{N \tau_{\textrm{comp}}}}
\end{align}

\begin{figure}
	\includegraphics[width=1.0\textwidth]{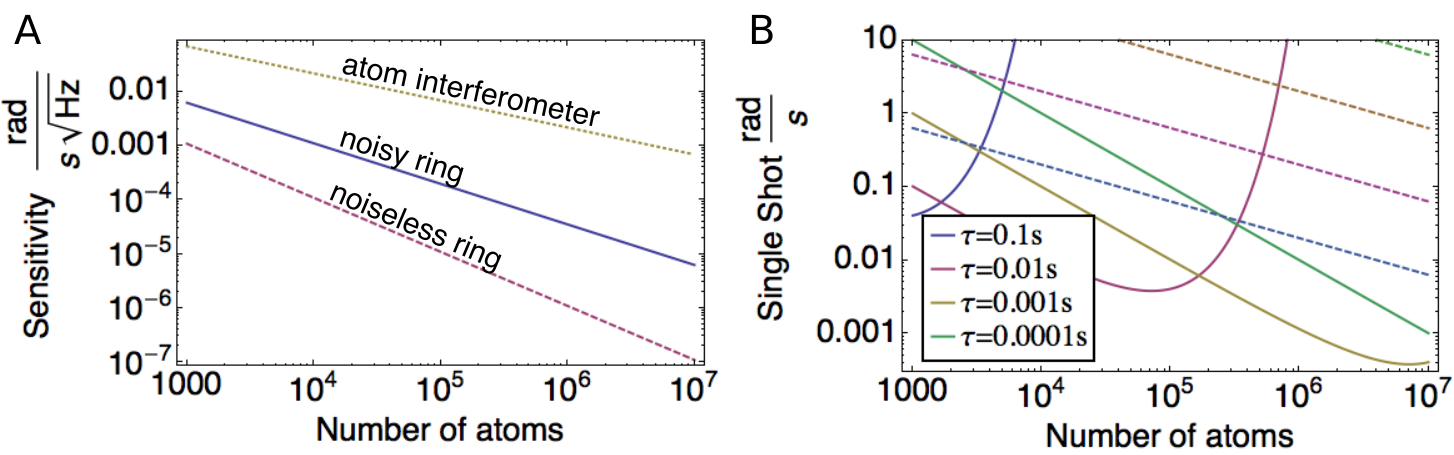}
	\caption{A. The sensitivity of our proposed gyroscope (solid) plotted on a log-log scale as a function of atom number. The dashed line represents the sensitivity for a noise-less system with an observation time of $\tau_{\textrm{comp}} = 0.838$ s. The dotted line represents an atom interferometer also running over $\tau_{\textrm{comp}}$. B. Solid (dashed) lines: the single-shot sensitivity for the noisy Luttinger (single atom) system for different observation times, $\tau$, as a function of atom number.}
	\label{fig:sens}
\end{figure}

From Fig. \ref{fig:sens}a we can see that the sensitivity is much better for the Luttinger ring system. The advantage over the single atom case increases with density, as expected from the atom number scaling. The single atom case shows scaling $\propto N^{-1/2}$ due to the shot-noise limit. Similarly, the noiseless Luttinger system shows the expected Heisenberg-like scaling ($\propto N^{-1}$). Fig. \ref{fig:sens}b. shows the single shot sensitivity for the noisy Luttinger system (solid lines), demonstrating the trade-offs between longer observation times and number fluctuation noise. For reference, the dashed lines are the values for a single atom system with the same observation times. Of course, these sensitivities obscure the difficulty in preparing the Luttinger system. Due to the nature of the gap between current states, the preparation time for superpositions with larger atom numbers grows quickly and the system will be less sensitive for repeated measurements.

A detailed analysis of the limitations on coherent superpositions in Luttinger liquids will be needed for a complete understanding of this type of gyroscope. Though we controlled the dominant dephasing mechanism by working slowly enough to avoid creating phonons, it is not clear how stable the superposition will be if particle loss is included. Simulations of small numbers ($<10$) of atoms suggest that the strongly repulsive regime considered here may be robust to particle loss but it is not clear if these results extend to many atoms \cite{Hallwood2010b}.

We have designed and characterized a new type of gyroscope, in which atomic interactions are employed to enhance sensitivity to small rotations.  This system uses a weak rotating laser barrier to create superpositions of persistent current states of strongly repulsive atoms.  The strong repulsion maintains the gap so that the process can be performed adiabatically, while the weak barrier allows the current states to retain most of their character to enable rotation sensing.  The scheme is sensitive to small rotation rates and shows favorable scaling, even in the presence of noise. In addition to rotation sensing, it is possible that creating these superpositions will have other interesting applications, e.g. for use as qubits \cite{Solenov2010c,Solenov2010b,Aghamalyan2015a} or to test inertial equivalence.

\section{Acknowledgements}
We thank L. Mathey, G. K. Campbell, and G. Zhu for insightful discussions and helpful feedback. Funding was provided by the NSF supported Physics Frontier Center at the JQI.

\bibliographystyle{apsrev4-1}
\bibliography{InteractingAtomicInterferometry2}

\end{document}